\documentclass[review]{elsarticle}

\usepackage{lineno,hyperref}

\journal{Journal of ASTP Templates}

\newcommand{\aap}{{Astron. Astrophys.}}
\newcommand{\apj}{{Astrophys. J.}}
\newcommand{\apjl}{{Astrophys. J. Lett.}}

\newcommand{\grl}{{Geophys. Res. Lett.}}

\newcommand{\mnras}{{Mon. Not. Roy. Astron. Soc.}}

\newcommand{\solphys}{{Solar Phys.}}

\newcommand{\ssr}{{Space Sci. Rev.}}
\newcommand{\cjaa}{{CHJAA！！}}
\newcommand{\apss}{{---！！}}

\begin{document}

\begin{frontmatter}

\title{Predicting solar surface large-scale magnetic field of Cycle 24}

\author[BUAA,NAOC]{Jie Jiang}
\cortext[mycorrespondingauthor]{Jie Jiang}
\ead{jiejiang@buaa.edu.cn}

\author[BUAA]{Jinbin Cao}



\address[BUAA]{School of Space and Environment, Beihang University, Beijing, China\corref{mycorrespondingauthor}}
\address[NAOC]{Key Laboratory of Solar Activity, National Astronomical Observatories, Chinese Academy of Sciences, Beijing 100012, China}

\begin{abstract}
  The Sun's surface field, especially the polar field, sets the boundary condition for the coronal and heliospheric magnetic fields,
  but also provides us insight into the dynamo process. The evolution of the polar fields results from the emergence and subsequent
  evolution of magnetic flux through the solar surface. In this paper we use a Monte Carlo approach to investigate the evolution of the
  fields during the decay phase of cycle 24. Our simulations include the emergence of flux through the solar surface with statistical
  properties derived from previous cycles. The well-calibrated surface flux transport model is used to follow the evolution
  of the large-scale field. We find the polar field can be well reproduced one year in advance using the observed synoptic magnetograms as
  the initial condition. The temporary variation of the polar field measured by Wilcox Solar Observatory (WSO), e.g., the strong decrease
  of the south polar field during 2016-2017 which is not shown by SDO/HMI and NSO/SOLIS data usually is not well reproduced.
  We suggest observational effects, such as the effect of the large gradient of the magnetic field around the southern
  polar cap and the low resolution of WSO might be responsible. The northern hemisphere polar field is predicted to increase during 2017.
  The southern polar field is predicted to be stable during 2017-2018. At the end of 2017, the magnetic field in two poles is predicted to be
  similar (although of opposite polarities). The expected value for the dipole moment around 2020 is 1.76$\pm$0.68 G and $2.11\pm0.69$ G based on the initial conditions from SDO/HMI and NSO/SOLIS synoptic magnetograms, respectively. It is comparable to that
  observed one at the end of Cycle 23 (about 1.6G based on SOHO/MDI).
\end{abstract}

\begin{keyword}
\texttt{Solar activity; Prediction; Solar magnetic fields; Solar polar fields}
\end{keyword}

\end{frontmatter}


\section{Introduction}
Prediction of short-term and long-term future levels of solar activity has found much interest in the literature since solar activity
is the driver of space weather, which has practical consequences for human activities in space. A reliable method of long-term prediction
could also provide constraints on dynamo models.

Sunspots are the most direct demonstration of solar activity. Most long-term predictions of solar activity concentrate on the prediction of
the time evolution of the sunspot number \cite{Hathaway1994,Li2005,Gholipour2005,Hiremath2008,Podladchikova2012} based on the extrapolation
method or the precursor methods, see \cite{Petrovay2010} for more details. Among the most reliable techniques are those that use geomagnetic
activity at or near the time of minimum as the precursors for future cycle amplitude \cite{Schatten1978,Schatten2005}. Geomagnetic activity
during the preceding cycle at or near the time of minimum corresponds to the axial dipole field \cite{Wang2009}. The good predictive abilities
of the geomagnetic indexes support the Babcock-Leighton (BL) type of solar dynamo, in which surface poloidal field is the source of the next
cycle strength. Since the source of the cycle strength in the BL dynamo is observable, it makes the dynamo-based solar cycle prediction
feasible. The dynamo-based predictions started around the end of cycle 23 by two groups \cite{Dikpati2006, Choudhuri2007,Jiang2007}, who gave
opposite predictions of Cycle 24 strength. See Section 4 of \cite{Karak2014} for the differences of two models. Although the dynamo-based
predictions attempted to model the time evolution of the magnetic field below and over the solar surface, they cannot predict the details of
the features of solar cycle, i.e, shapes of the solar cycle and the time evolution of the polar field.

Apart for their role in the solar dynamo, the polar field are of great importance in determining the global structure of the corona, e.g. \cite{Hayashi2008, Hayashi2013}, the heliospheric magnetic fields, the propagation of galactic cosmic rays throughout the heliosphere and so on. However, owing to foreshortening effects at the solar limb, it is hard to accurately measure the evolution of the polar fields. Determinations of the polar fields are
further complicated by the variable $B_0$ angle of the Sun's rotation axis with respect to the ecliptic \cite{Petrie2015,Wang2016}.

Magnetic flux generated by the dynamo process in the interior emerges at the solar surface in the form of bipolar magnetic regions (BMRs)
with a preferred tilt of the axis joining the two polarities with respect to the equator. The emerged flux is then transported and dispersed
over the solar surface due to systematic and turbulent motions. When magnetic flux elements of opposite polarity comes into contact,
the features cancel, removing equal amounts of flux of each sign. Because of the systematic tendency of the tilt angle, a net flux is
transported across the equator during each cycle. This leaves a net surplus of following flux in each hemisphere which is carried
poleward by the surface meridional flow. The accumulation of these remnants of BMRs eventually neutralizes, reverses, and builds up
the polar field for the next cycle  as the sunspot cycle progresses. The whole process can be well simulated by the Surface Flux
Transport (SFT) models \cite{Wang1989, Ballegooijen1998, Baumann2004, Yeates2007}. The model show remarkable success in reproducing
the evolution of the Sun's large-scale field over surface, although some differences including the transport parameters and the
methods to treat the flux source term, are used by different authors.

The success of the SFT model in reproducing the large-scale field demonstrates its functionality and potential for predicting the
large-scale field evolution \cite{Schrijver2003}. Based on the statistical properties of solar cycle, we may predict the sunspot
emergence. With the help of the SFT model, people can get the possible large-scale field evolution over the surface, including the
polar field and the axial dipole moment a few years in advance. In this study, we aim to predict the sunspot emergence, polar field strengths,
the appearance of the magnetic butterfly diagram during 2017-2020, and the strength of cycle 25. The differences from \cite{Cameron2016a}
concentrate on three aspects. Firstly, we have slightly improved the empirically derived statistics of the solar cycle variation of
the sunspot group emergence. Secondly, we include predictions for the polar field strength and the magnetic butterfly for the rest of cycle 24.
Thirdly, we updated the prediction of cycle 25 based on the assimilation of the most recent data into the model.

The paper is organized as follows. In Section 2, we present our improved descriptio of sunspot emergence properties based on the solar cycle
properties. The details about the SFT model are given in Section 3. In Section 4, we give our predicted results about the large-scale field
evolution during 2017-2020 and the possible strength of cycle 25.

\section{Prediction of sunspot emergence}
\label{sec:source}
A main ingredient of the SFT model is the emergence of bipolar magnetic fields.  In this section we present an improved description
of the sunspot group emergence, which includes the number, location, area, and tilt angle, and the the empirical statistical properties
were used to derive them.

\begin{figure}[!htp]
\includegraphics[scale=0.4]{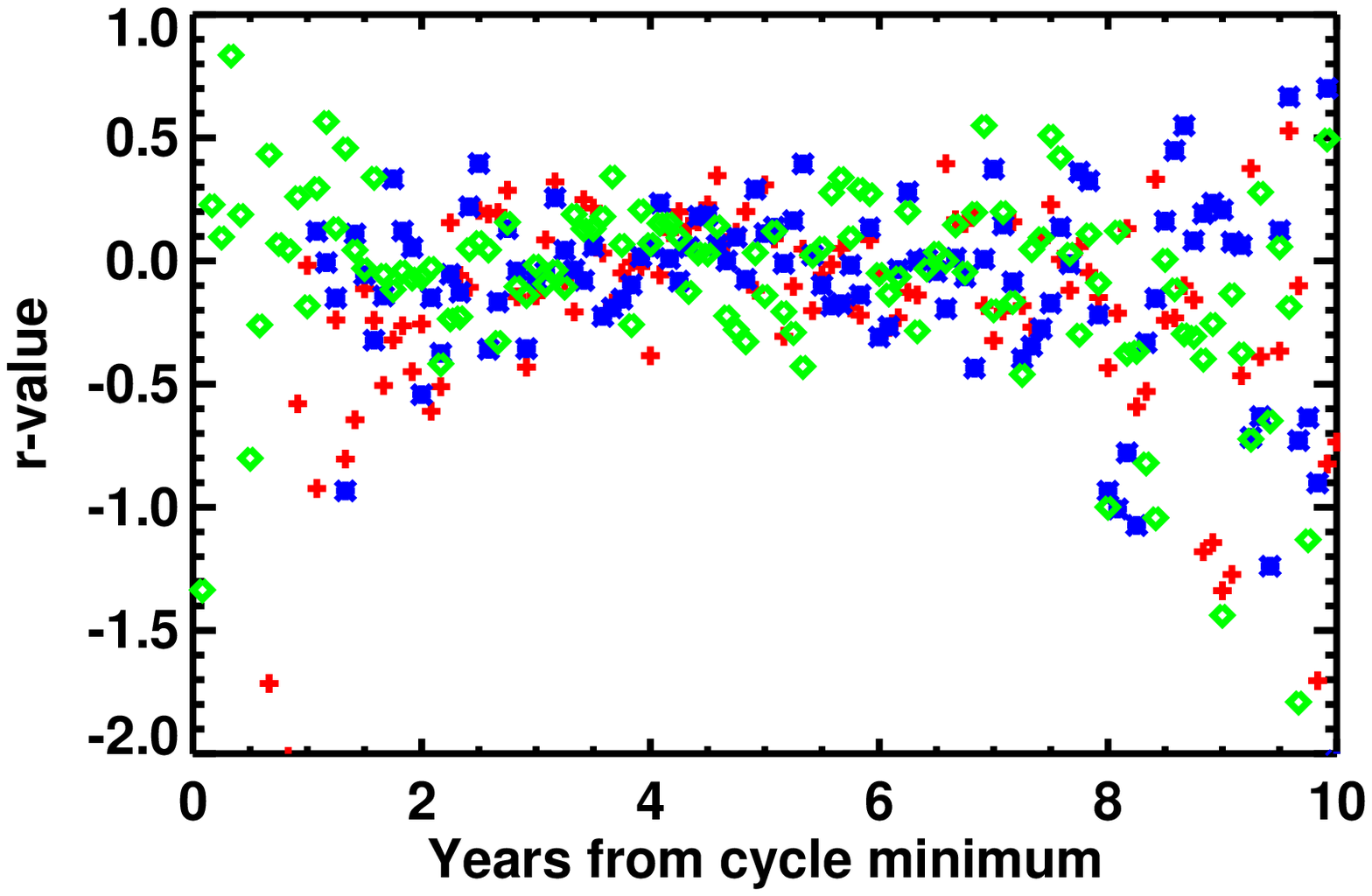}
\includegraphics[scale=0.4]{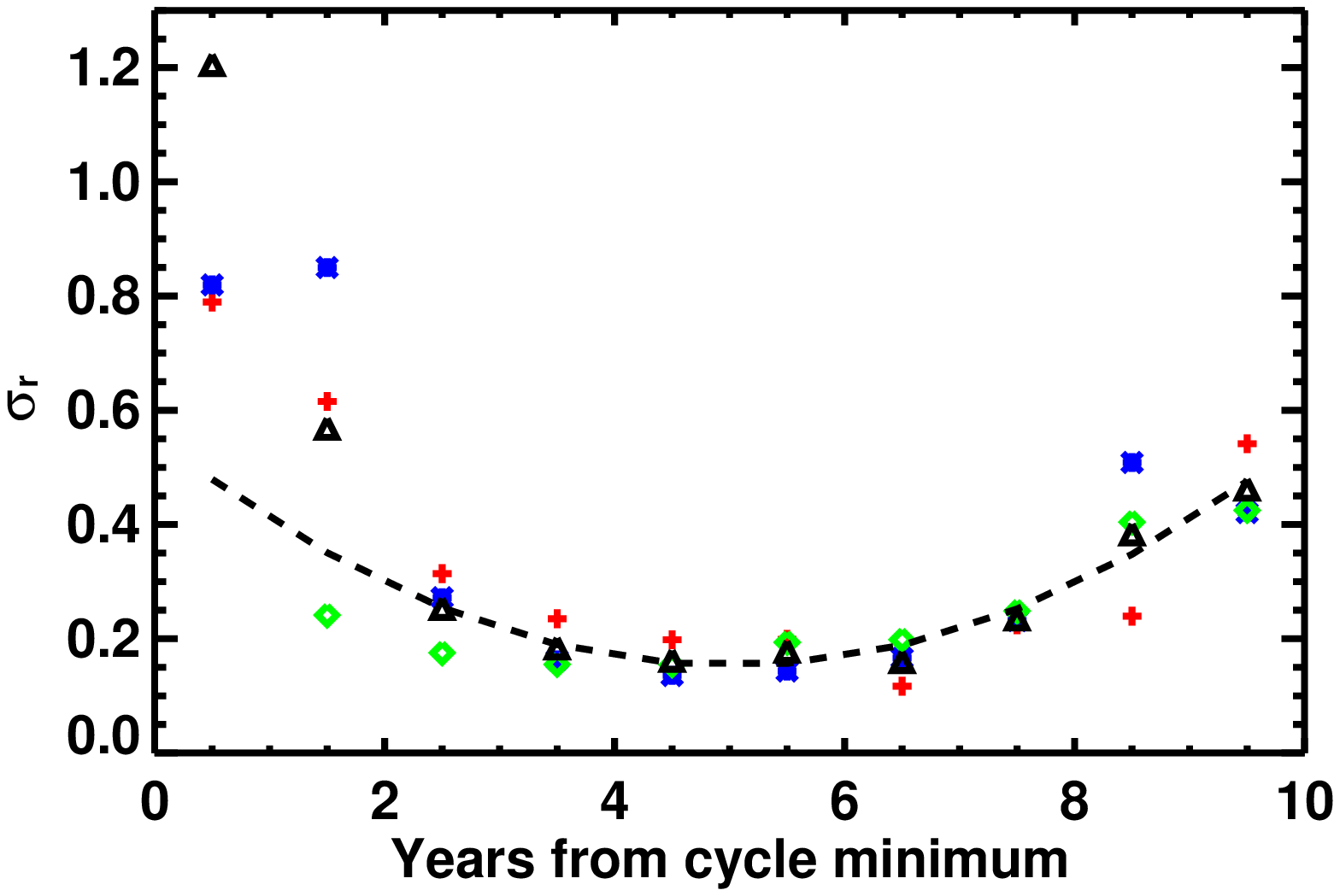}
\caption{Left panel: Ratio between the difference $\Delta f(t)$ and $f(t)$ since the starts of Cycle 21 in blue, Cycle 22 in green,
  and Cycle 23 in red, where $f(t)$ is the fitted function of the solar cycle and $\Delta f(t)$ is the random scatter of the sunspot number.
  Right panel: Standard deviation ($\sigma_r$) over 1 yr bins of ratio for different cycles in colors. The black triangles denote the
  averages over the 3 cycles. Omitting the first 2 yrs due to the cycle overlap, the time evolution of $\sigma_r$ over the cycle is fitted
  by $\sigma_r=0.55-0.16t-0.016t^2$ in dashed black curve.}
\label{fig:sn_scatter}
\end{figure}

Comparing with \cite{Cameron2016a}, we made the following improvement in deriving the time evolution of the BMR emergence.
Firstly, the new version of the monthly sunspot number ($R$) \citep{Clette2014} \footnote{http://www.sidc.be/silso/datafiles} was used here.
Based on the sunspot data since 1878 onwards, the number of BMRs emerging per month was taken to be equal to $R_G=0.24R$. The functional
form of $R_G$ is $R_G=f(t)+\Delta f(t)$, where $f(t)$ was given by Equation (1) of \cite{Hathaway1994},
It is $f(t)=a(t-t_0)^3/\{exp[(t-t_0)^2/b^2]-c\}$, where $b(a)=27.12+25.15/(a\times10^3)^{1/4}$ and $c=0.71$.
The values for parameters \emph{a} (amplitude) and \emph{$t_0$} (starting time) to get $f(t)$ are 0.0018 and 2008.98. $\Delta f(t)$
denotes the random scatter of the time evolution of the sunspot number. The left panel of Figure \ref{fig:sn_scatter} shows the evolution
of the ratio (r) between $\Delta f(t)$ and $f(t)$ for cycles 21-23. The right panel shows the evolution of the standard deviation of the
ratio ($\sigma_r$) with one year bin size for the three different cycles. The symbol of black triangle is the averaged values. The fitted
function of the standard deviation of the ratio excluding the first two year when there is large scatters due to the cycle overlap is
$\sigma_r=0.55-0.16t-0.016t^2$.

The latitudinal distribution and the mean tilt angle were studied by \cite{Jiang2011a}. Since the new sunspot record was used here,
  we recalculated the relation between the mean latitudes $\lambda_n$ and cycle strength $S_n$ for different cycle $n$ using the method
  suggested by \cite{Jiang2011a}. The latitudinal distribution is $\lambda_n=12.2+0.015S_n$. For the scatter of latitude distribution
(standard deviation $\sigma_n^{i}$ for cycle $n$ at $i$th phase of the cycle), we excluded points deviating from the mean by more than
2$\sigma_n^i$ in \cite{Jiang2011a, Cameron2016a}, which caused the sharp boundary of the butterfly diagram compared with the observations.
Here we only exclude points deviating from the mean by more than 2.2$\sigma_n^i$ on the equatorward side in order to better reproduce the
butterfly diagram. The mean tilt angle, $\alpha_n$, obeys $\alpha_n=T_n\sqrt{|\lambda}|$. The relation between $T_n$ and $S_n$ is
taken as $T_n=1.72-0.0022S_n$. For the scatter of the tilt angle, we used the empirical relation which depends on sunspot umbral area,
i.e., Equation (1) of \cite{Jiang2014}. The resulting tilt angle is multiplied by a factor 0.7 to include the effect of inflow towards
the activity belts \cite{Cameron2010, Jiang2010ApJ, Martin-Belda2016, Martin-Belda2017}. The area distribution is based on the Equations
(12)-(14) of \cite{Jiang2011a}. The BMRs have a random distribution in longitudes.

\begin{figure}[!htp]
\includegraphics[scale=0.35]{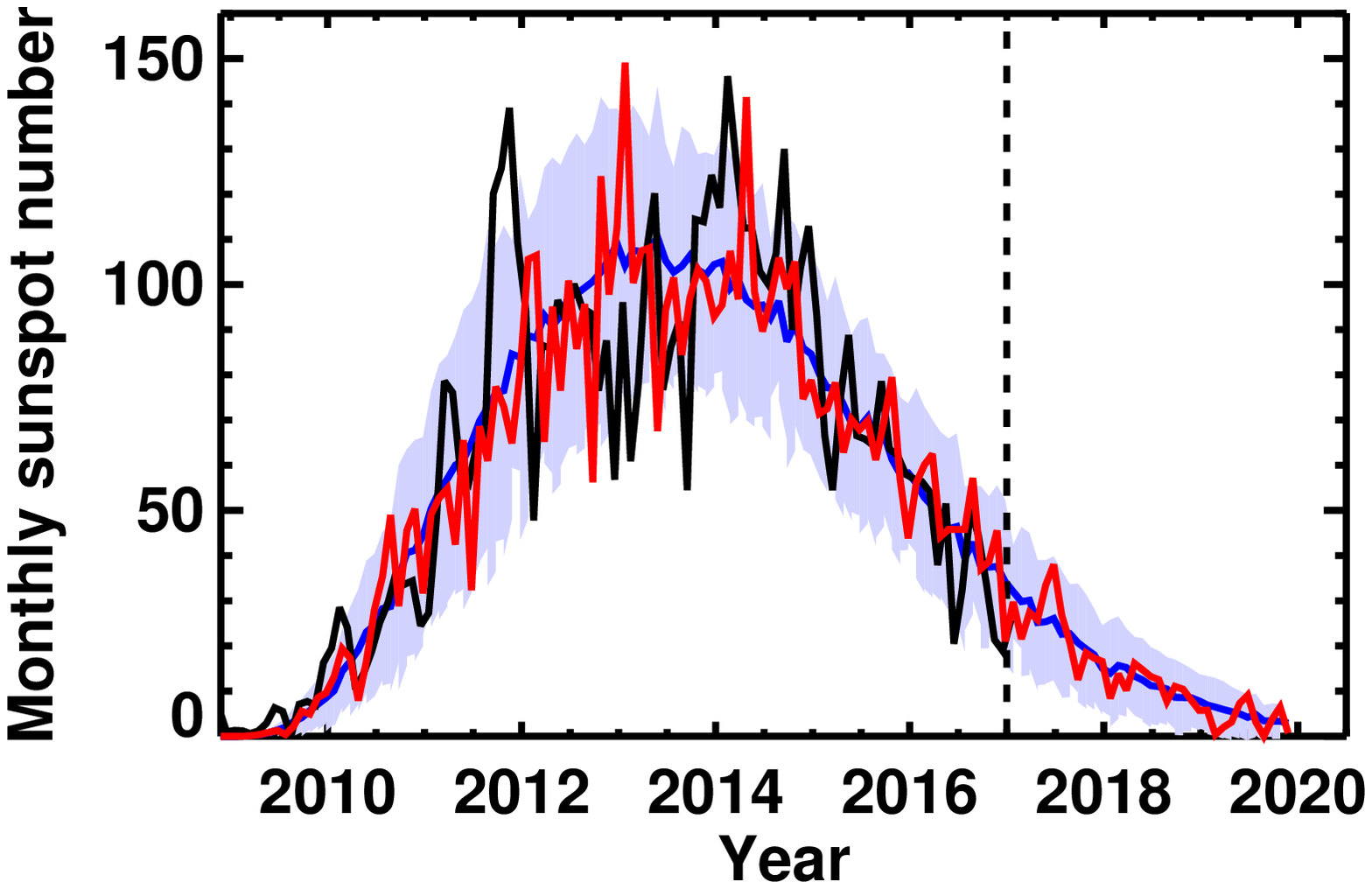}
\includegraphics[scale=0.35]{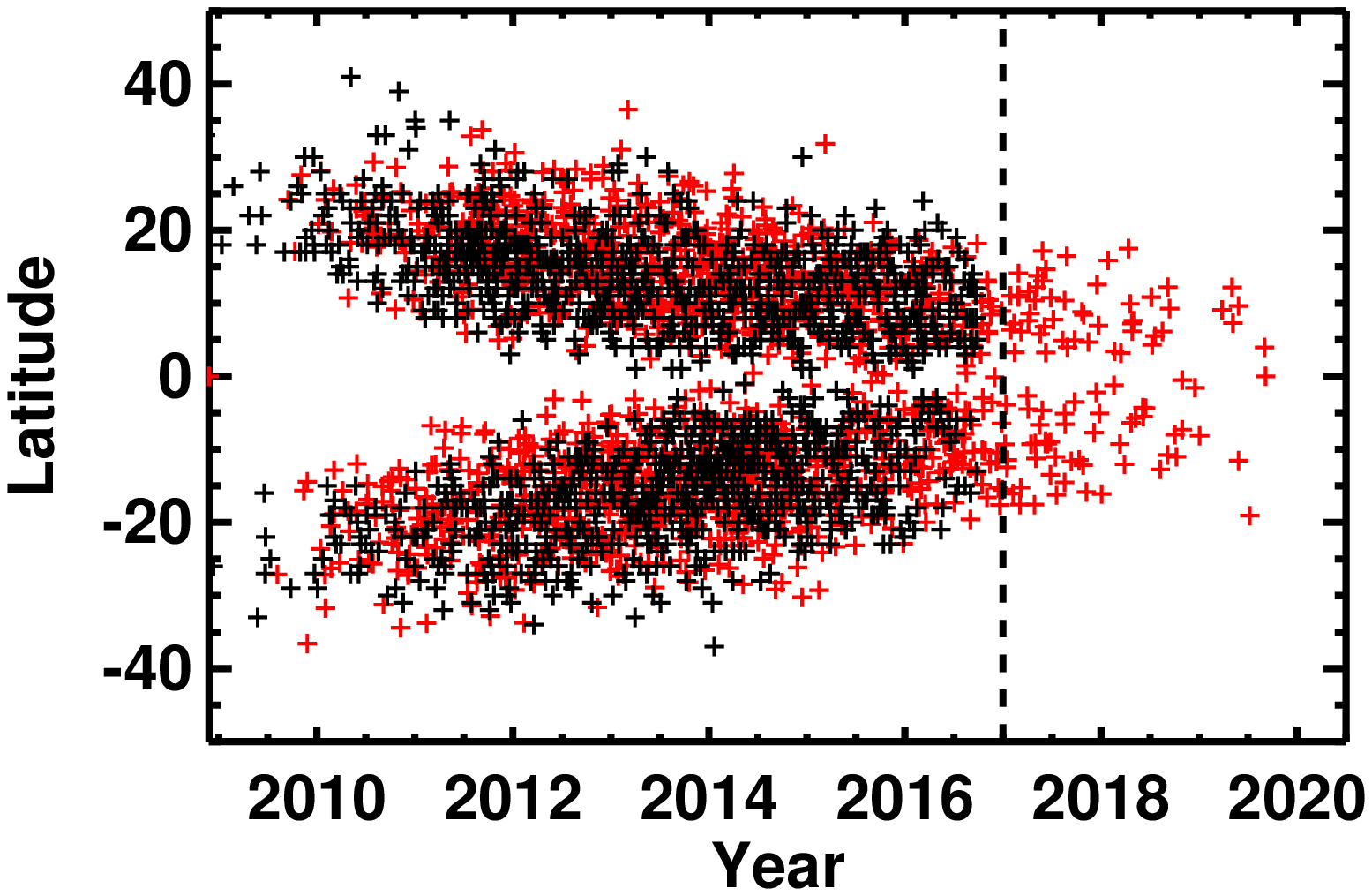}
\caption{Example of a realization of Cycle 24 with Monte Carlo simulation of random sunspot emergence. Left panel: monthly group sunspot
  number (black curve: observed; red curve: one random realization). The blue curve and the shading show the average and the $\pm2\sigma$
  variation of 50 random realizations. Right panel: butterfly diagram of emerging sunspot groups (Black crosses: observed sunspot groups;
  red crosses: one realization of random sources, corresponding to the red curve in the left panel.)}
\label{fig:sn_btf_pre}
\end{figure}

We made a Monte Carlo analysis using 50 realizations of sunspot emergence generated from the standard deviation of time evolution of
sunspot group number. Figure \ref{fig:sn_btf_pre} shows the comparison of the actual monthly sunspot number (left panel) and of the butterfly
diagram (right panel) with the random realizations based on the statistical relations. The blue shading indicates the $\pm2\sigma$ variation
of 50 random realizations. The observed monthly sunspot number are almost within the shading. The red curve is an example for one realization
of random source of Cycle 24.

Comparing to Figure 1 of \cite{Cameron2016a}, Figure \ref{fig:sn_btf_pre} is closer to the observational case. There is a larger scatter
for the sunspot number during the decay phase, which plays important roles in the polar field strength at the end of the cycle since
they have low latitude distribution \cite{Cameron2013,Jiang2014}. The butterfly diagram is more similar to the observed one,
especially near the boundaries of the butterfly wings.

\section{Surface flux transport modelling}
\subsection{Surface flux transport model}
With the sunspot group emergences as the source of the surface magnetic flux, the SFT model can be used to study the
evolution of the magnetic field over the surface. The model treats the evolution of the radial component of the large-scale magnetic
field $B$ at the solar surface resulting from passive transport by convection (treated as a diffusivity), differential rotation $\Omega$,
and meridional flow $\upsilon$. The corresponding equation is
\begin{eqnarray}
\frac{\partial B}{\partial t}=&-&\Omega(\theta)\frac{\partial
B}{\partial \phi}-\frac{1}{R_\odot\sin\theta}
\frac{\partial}{\partial\theta}\left[\upsilon(\theta)B\sin\theta\right]\nonumber \\
&+&\frac{\eta}{R_\odot^{2}}\left[\frac{1}{\sin\theta}\frac{\partial}{\partial\theta}
\left(\sin\theta\frac{\partial B}{\partial\theta}\right)+
\frac{1}{\sin^{2}\theta}\frac{\partial^{2}B}{\partial\phi^2}\right]\nonumber\\
&+&S(\theta,\phi,t),
\end{eqnarray}
where $\theta$ and $\phi$ are heliographic colatitude and longitude, respectively. The magnetic diffusivity, $\eta$ describes the
random walk of the magnetic flux elements as transported by supergranulation flows. The source term, $S(\theta,\phi,t)$, descibes the
emergence of magnetic flux at the solar surface. The time evolution of $S$ is obtained using the randomly realized sunspot emergence in
Section \ref{sec:source}.

We take the same flux transport parameters, i.e., $\Omega(\theta)$, $\upsilon(\theta)$ and $\eta$ as \cite{Cameron2016a}.
There is no evidence for the time variation of the supergranulation flows \cite{Meunier2007,Rieutord2010}, and hence of the magnetic
diffusivity. The amplitude of the time variation of the differential rotation, i.e., the torsional oscillation is in one thousandth of
the overall rotation and thus has negligible effects. The cycle-phase dependence of the meridional flow \cite{Chou2001, Hathaway2010}
is mostly due to localized inflows into active regions and is here included only by multiplying the tilt angles a factor 0.7 \cite{Cameron2010}.

\subsection{Initial conditions}
We used the first synoptic magnetogram of 2017, corresponding to Carrington rotations CR2185, from NSO/SOLIS and SDO/HMI as the initial conditions
for the SFT simulations. Synoptic magnetograms of CR2173 were also used to test our prediction method.  The magnetogram data used were
reduced to a resolution of 1 degree in latitude and longitude. For the HMI magnetograms, the well-observed polar data obtained in each spring
or fall are interpolated to estimate the radial field above 75$^\circ$ latitude at any given time and the smoothed, interpolated values are
used to fill in the regions with data missing due to the unfavorable viewing angle \cite{Sun2011}. The projection effect due to the
  $B_0$-angle effect was also removed. For the NSO synoptic maps, the unobserved polar fields were filled in using a cubic-polynomial surface
  fit to the observed field at neighboring latitudes. The project effect of other latitudes was not corrected. Hence the level of uncertainties
  of the polar field in the SDO/HMI synoptic maps is smaller than that in NSO/SOLIS data.
The 50 random realizations of sunspot emergence during 2017 onwards are taken as realizations of possible sunspot emergences.

We use the same method as \cite{Cameron2016a} to estimate the contribution of the measurement error due to net flux density in the initial
magnetogram to the uncertainty of the polar field and the axial dipole moment. The final error is the quadratical summation of the error by
the uncertainty due to scatter in the properties of the BMR source and the error by the magnetogram.

\section{Results}
\label{sec:results}
\subsection{Prediction tests: Large-scale field evolution during 2016-2017}
It takes about one year for the sunspot emergence at the activity belts to be transported to the latitudes above 55 degree.
So we expect the random sunspot emergence at activity belts has negligible effects on the mean flux density over polar caps
during the first year. We first test the predictive abilities of this model to generate the large-scale field, including polar field,
axial dipole field and butterfly diagram by attempting to reproduce these parameters during 2016-2017.

The longitudinally averaged radial field as a function of colatitude and time, which corresponds to a ``magnetic butterfly diagram"
is defined as
\begin{equation}
\left\langle B
\right\rangle(\theta,t)=\frac{1}{2\pi}\int_{0^\circ}^{360^\circ}B(\theta,\phi,t)d\phi.
\label{eq:bavg}
\end{equation}
The WSO has measured the polar field since 1976. Due to the long history and the homogeneity of the dataset, they have been widely used by the communities. The line-of-sight field between about $55^\circ$ and the poles was measured everyday. In order to compare with the WSO polar field evolution, we define the LOS polar field for the north pole as
\begin{equation}
  B_{LOS} = \int_0^{35^\circ} \left\langle B\right\rangle(\theta,t)
              \sin^{2}\theta d\theta\left/\int_{0}^{35^\circ}\sin\theta d\theta\right.,
\label{eq:polar1}
\end{equation}
and analogous for the LOS southern polar field. NSO provides the daily photospheric radial polar field measurements since
October 2006  \footnote{http://solis.nso.edu/0/vsm/vsm\_plrfield.html}. Three separate bands of latitude are considered for each
hemisphere: $\pm60^\circ$ to $\pm75^\circ$, $\pm60^\circ$ to $\pm70^\circ$, $\pm65^\circ$ to $\pm75^\circ$. Due to their substantial
increase of the noise of the polar field, higher latitudes were not included. To make a definition consistent with NSO polar fields
within $\pm60^\circ$ to $\pm75^\circ$, we define the radial polar field as
\begin{equation}
  B_{r} = \int_{15^\circ}^{30^\circ} \left\langle B\right\rangle(\theta,t)
              \sin\theta d\theta\left/\int_{15^\circ}^{30^\circ}\sin\theta d\theta\right.,
\label{eq:polar2}
\end{equation}
and an analogous quantity for the radial southern polar field.

The definitions of the mean polar field strength involve the selected latitude bands. The original magnetogram pixels have to be
divided into subpixels in order to determine the average field strength in the bands. If the spatial resolution of the magnetograms
is low, then the large gradient of the field around the edge of the polar cap, leads to errors in the interpolation which turns out to
be a substantial source of inaccuracy in the determination of the polar field. This will be demonstrated by the WSO observations in the
subsequent section. The axial dipole moment which is defined based on the global field, thus turns out to be a better metric for the
description of the global large-scale field evolution \cite{Upton2014}. It is calculated in the following form
\begin{equation}
D(t)=\frac{3}{2} \int_0^{180^\circ} \left\langle B\right\rangle(\theta,t)
              \cos\theta\sin\theta d\theta
\end{equation}
\label{eq:dipole}

\begin{figure}[!htp]
\includegraphics[scale=0.35]{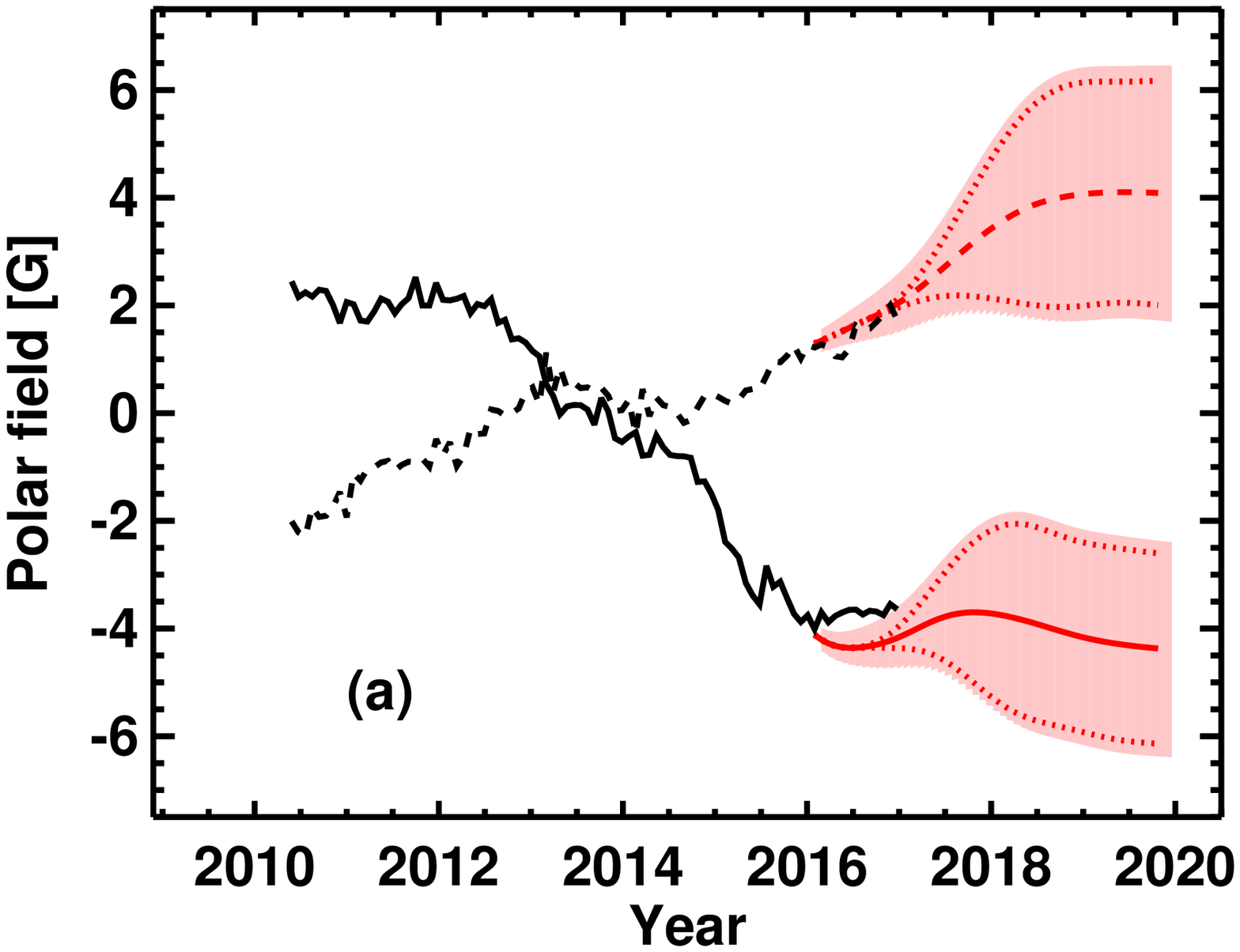}
\includegraphics[scale=0.35]{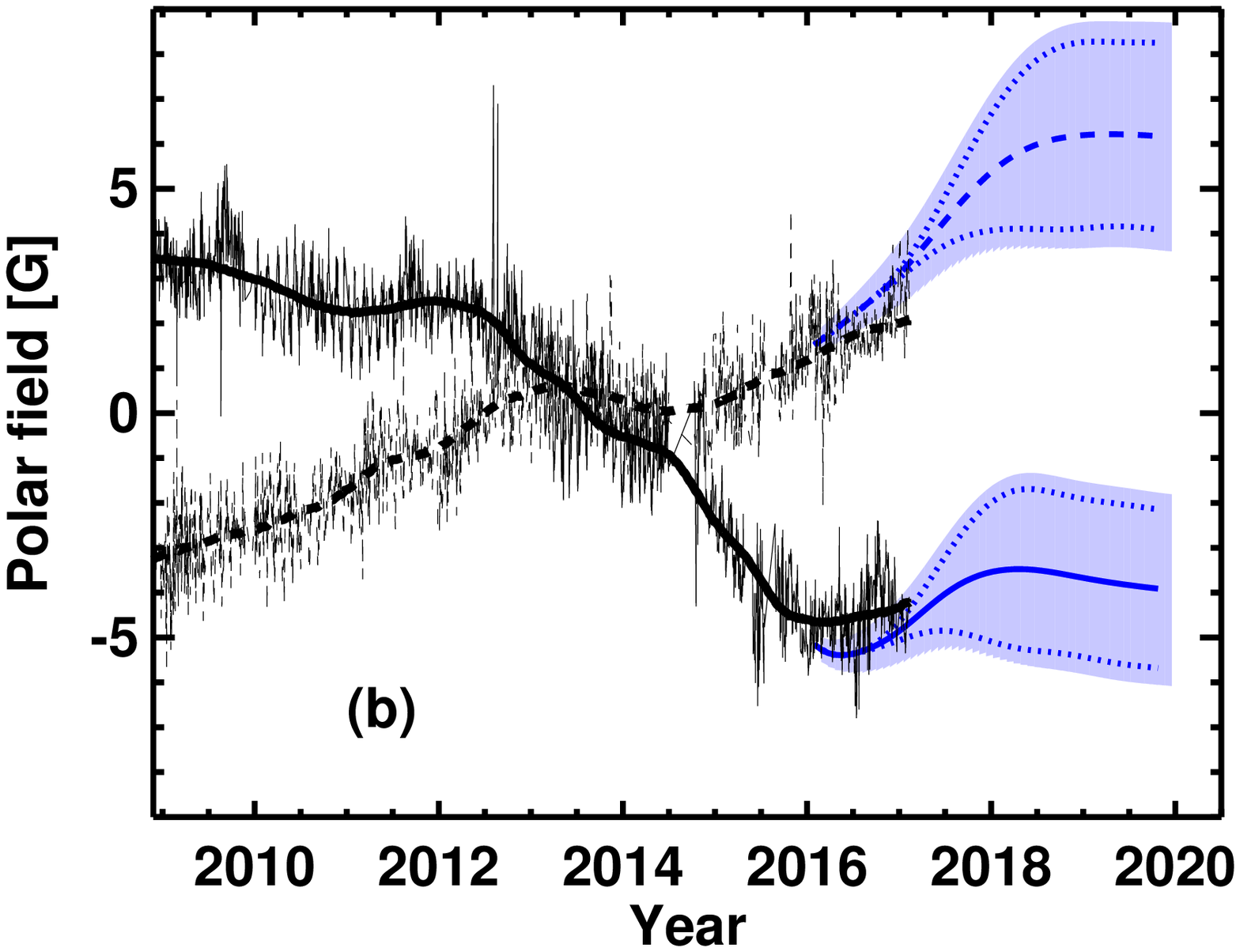}
\includegraphics[scale=0.35]{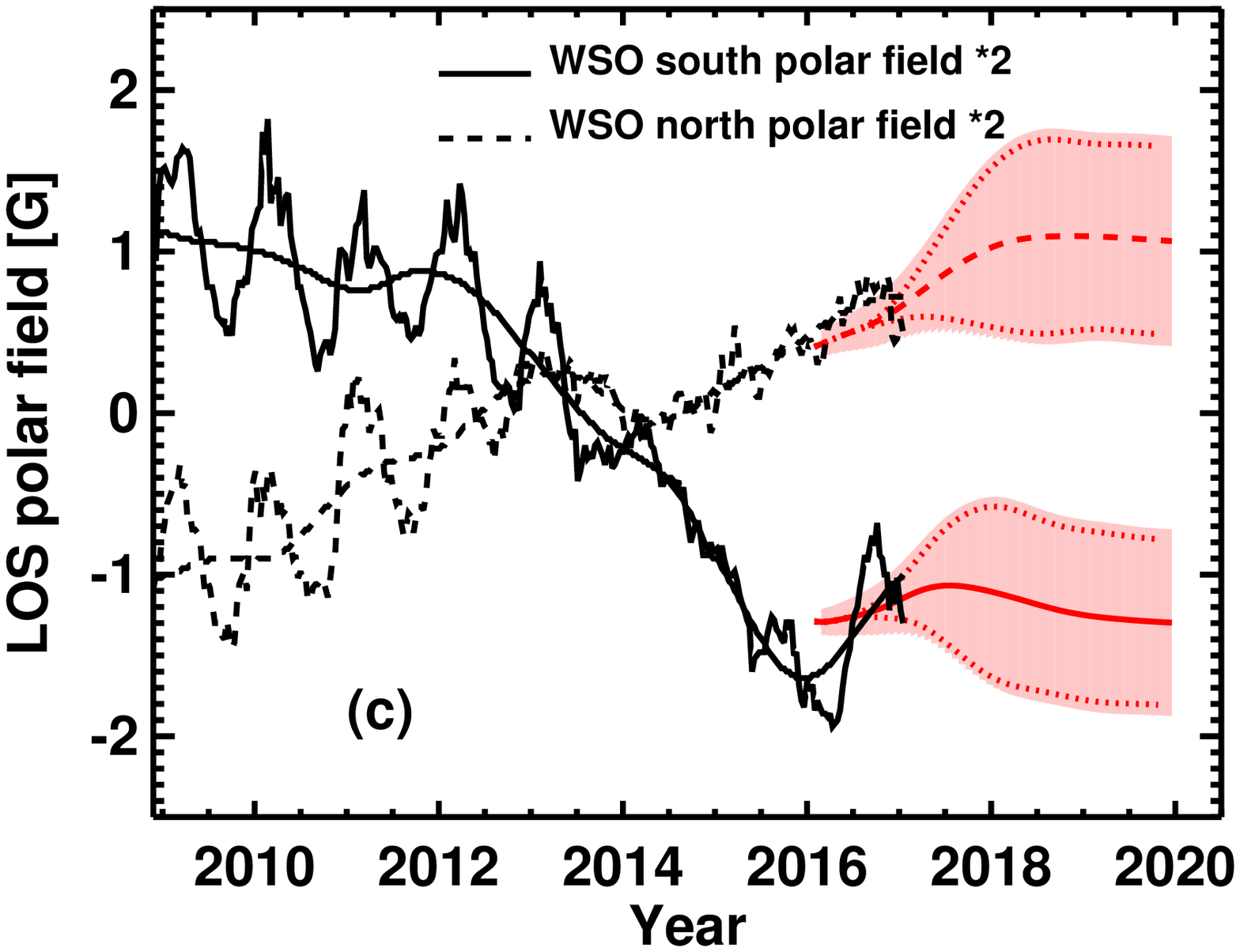}
\includegraphics[scale=0.35]{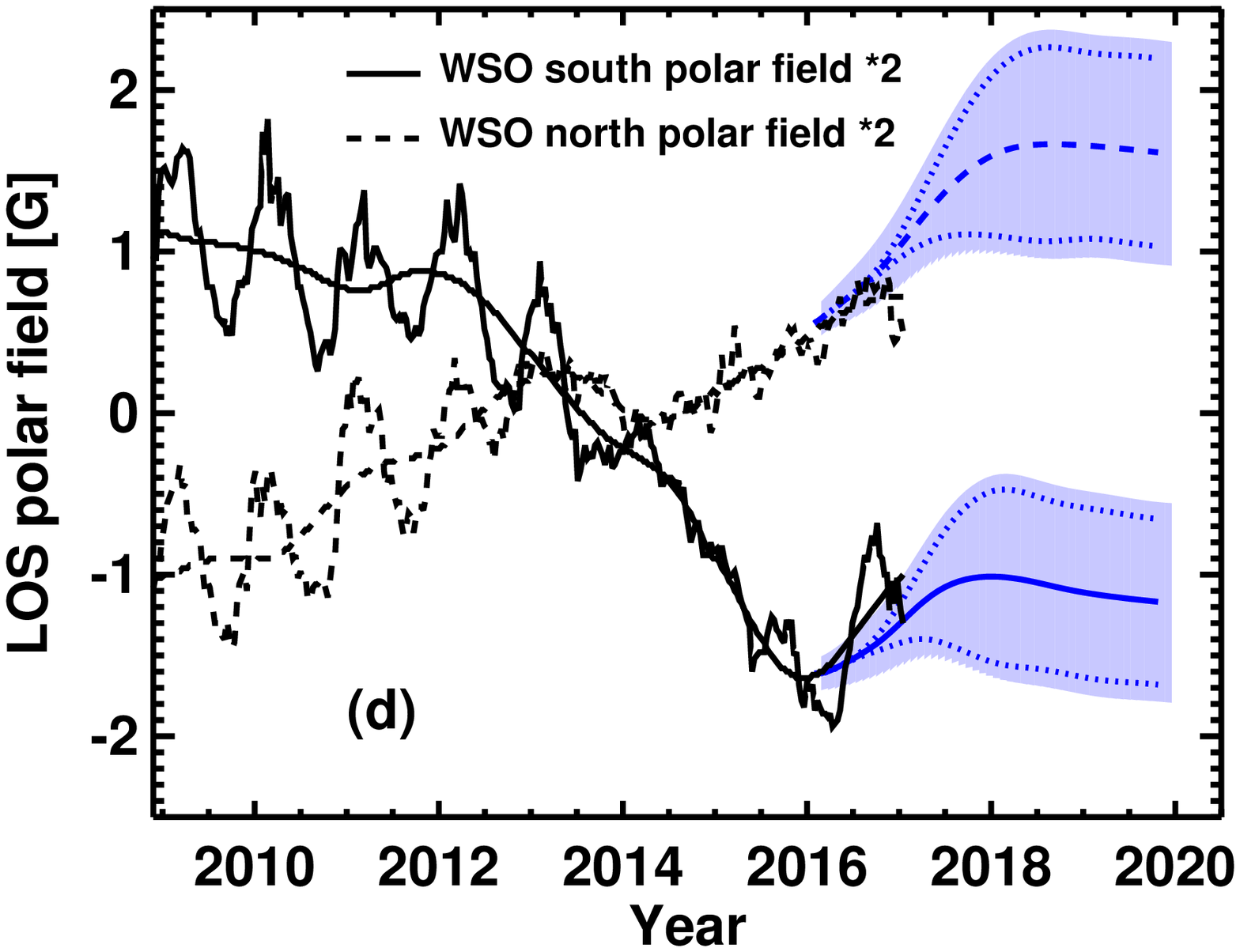}
\caption{Comparison of the polar field evolution between observations and predictions -- Solid curves: south polar field;
  Dashed curves: north polar field; red curves: prediction using the HMI synoptic magnetogram CR2173 as the initial condition;
  blue curve: prediction using the NSO/SOLIS synoptic magnetogram CR2173 as the initial condition. The shaded region gives the
  $\pm2\sigma$ variation of 50 random realizations. The curves in the middle of the shading regions denote the average of the 50
  random realizations. The dotted curves give the 2$\sigma$ range for the intrinsic solar contribution (source scatter).
  Panel (a): time evolution based on HMI observations during 2010-2017 in black and the predicted one from 2016.
  Panel (b): time evolution based on NSO/SOLIS observations during 2008.9-2017 in gray and the corresponding
  filtered data in thick black. Panel (c): time evolution of the LOS polar field based on WSO observations during 2008.9-2017 in gray
  and the predicted one using the HMI synoptic magnetogram as the initial condition. The difference between Panel (c) and Panel (d) is
  only the initial condition.}
\label{fig:pf2016}
\end{figure}

Figure \ref{fig:pf2016} shows the comparison between the observed polar field and the predicted results. The Black curves in
Panel (a) are calculated based on the SDO/HMI synoptic maps, the red curves are the predicted polar field evolution using the
SDO/HMI synoptic magnetogram of CR2173 as the initial condition. The shading indicates the $\pm2\sigma$ variation of 50 random realizations.
The random realizations of the sunspot emergence in the activity belts cause the uncertainties of the polar field one year later, i.e.,
from 2017 onwards. The predicted polar field in the norther hemisphere is consistent with the observed one. The polar field in the
southern hemisphere keeps on increasing after the reversal. It stops increasing and starts a slight decrease from 2016. The predicted
curve shows slightly delay of the decrease, which causes the averaged polar field 12\% stronger than the observed one.
But  they are within 3$\sigma$ range of the uncertainty due to the initial condition.

The thin black curves in Panel (b) of Figure \ref{fig:pf2016} are the time series of NSO/SOLIS polar mean radial field.
The thick black curves are the filtered time series where the strong annual modulation due to the variable $B_0$ angle
has been removed. The significant effects of the $B_0$ angle on the measured polar field are shown here. The blue curves show the
predicted polar field evolution using the NSO/SOLIS synoptic magnetogram of CR2173 as the initial condition. The variable $B_0$-angle
effect is not removed from the initial magnetograms. So we see that the initial polar field, which is stronger than the filtered one,
causes the subsequent predicted polar field stronger than the observed filtered values.  But they are within the range of the unfiltered data.

Since the WSO LOS polar field usually is taken as a reference in many studies, we compare the predicted polar field calculated
based on Equation \ref{eq:polar1} with WSO results. The thin and thick black curves in Panels (c) and (d) are the measured time
evolution of the WSO LOS polar field and the filtered results. WSO polar field shows a significant decrease after 2016.
The red and blue cures in Panel (C) and Panels (D) are based on the initial conditions of SDO/HMI and NSO/SOLIS CR2173
synoptic magnetograms, respectively. They are calculated by Equation \ref{eq:polar1}. In both panels, the predicted north
polar field matches the observed one well, but the prediction  fails to produce the strong decrease of the south polar field.

\begin{figure}[!htp]
\includegraphics[scale=0.35]{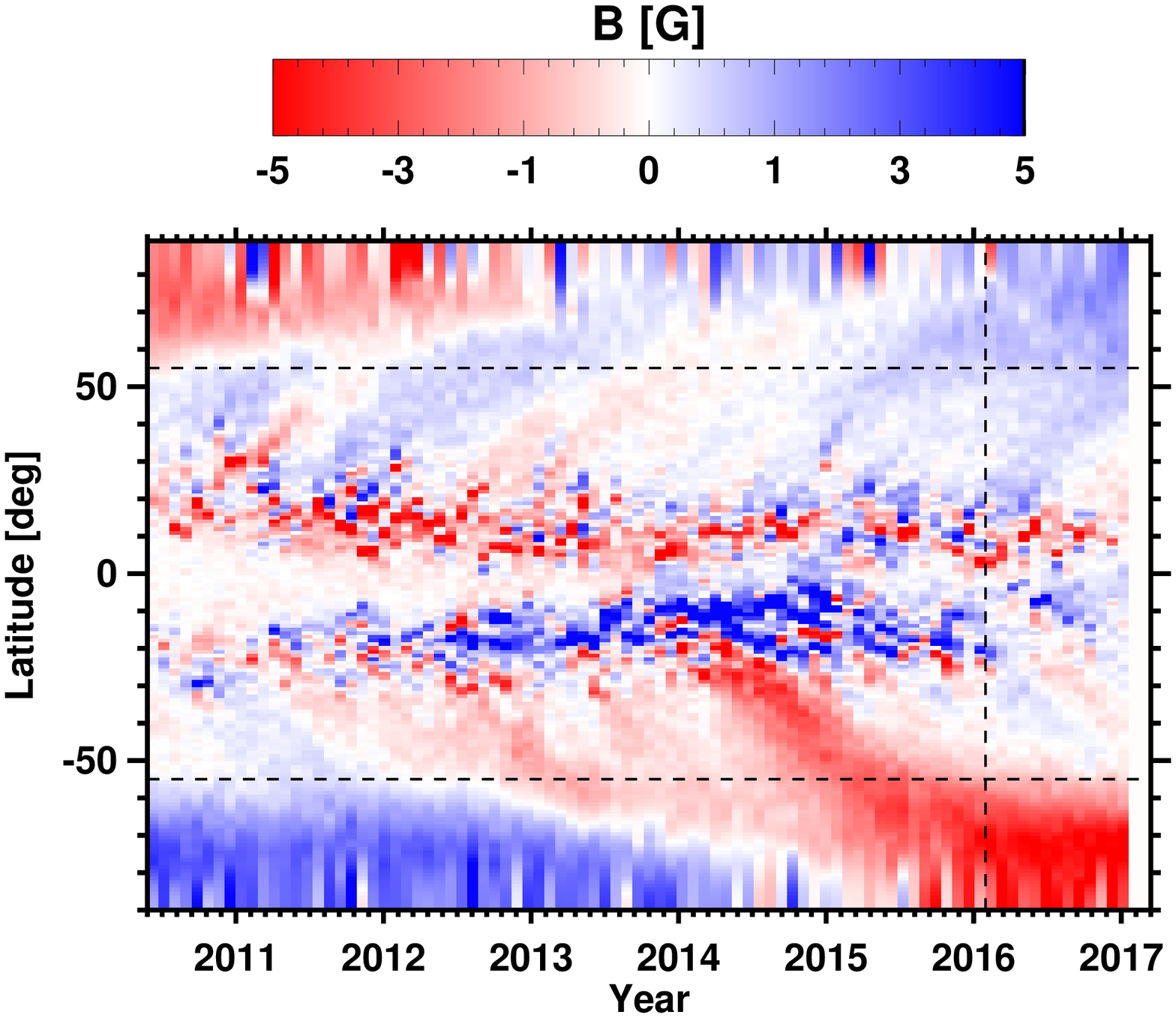}
\includegraphics[scale=0.35]{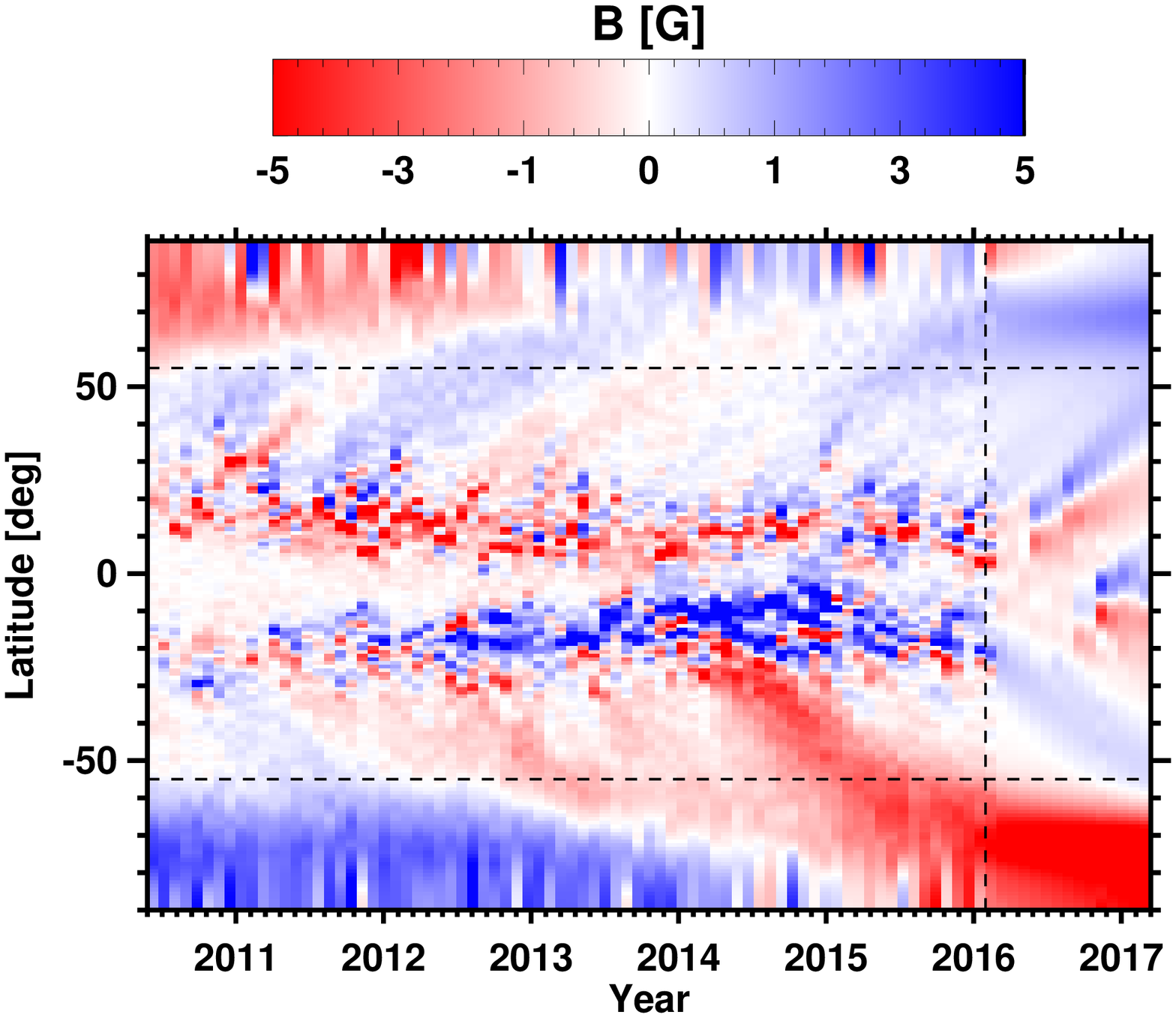}
\caption{Time-latitude diagrams of the longitudinally averaged radial magnetic field at the solar surface. Left panel: observation
  (SDO/HMI synoptic magnetic field data) during 2010-2017; Right panel: SDO/HMI synoptic magnetic field data during 2010-2016
  combined with the simulated result from one random realization of the flux source during 2016-2017.}
\label{fig:MagtfHMI}
\end{figure}

Figure \ref{fig:MagtfHMI} shows the magnetic butterfly diagram based on the SDO/HMI synoptic magnetograms.
The signal between 2016 and 2017 in the right panel is from the simulation of on one random realization.
The correlation between the observed and predicted behaviors during 2016-2017 is statistically significant correlated,
with a correlation coefficient 0.89. In the northern polar cap above 55 degrees, the large-scale diffusive positive field
is uniformly distributed over the polar region. In contrast, the structure is more complicated in the southern hemisphere.
The strong negative poleward plume starting from February 2014 reached the boundary of the pole cape, i.e. 55 degree in
the middle of 2015. Some of the positive flux from the leading polarity of sunspot groups reaches the boundary of the polar
region about half year later. An abnormal emergence (abnormal polarity orientation in the north/south direction with large area),
AR12403 on the solar disk during August 20-27, 2015 (later denoted as AR12418 and AR12431) also contributed to the net positive
flux transported to the southern polar cap. Hence the south polar field stops increasing, even shows the weak decrease.

Projection effects, the $B_0$ angle and the intrinsic weakness of mean flux density near the poles
  present difficulties to accurately measuring the polar field. To accurately resolve the boundaries of the
  latitude bands, the methods suggested by \cite{Bertello2014} were adopted. This divides the NSO/SOLIS original
  magnetogram pixels into subpixels to get the polar field evolution. Additionally the effect of the $B_0$ angle effect
  has been carefully removed, see the technical Report NSO/NISP-2015-002 for more details by Bertello \& Marble. In contrast,
  WSO measures the polar field in the polemost $175^{''}$ square apertures each day, north and south \cite{Svalgaard1978}.
  The latitude of the equatorward limit of that aperture is 55$^\circ$ on average. The $B_0$ angle effect is eliminated by 20nHz
  low-pass filter. When there are large gradients of the magnetic field around the polar region, the deviation from the average
  latitudes can introduce a large error to the measurement. For the southern hemisphere during 2016, based on our simulation,
  the measurement by WSO is expected to be about 0.4G lower if the polar field is defined as beginning at 50$^\circ$ rather than 55$^\circ$.
  For the northern hemisphere, where the field is almost uniformly distributed, a 5$^\circ$ change in the definition  has
  a negligible effect on the measurement.  We propose that the strong gradients on the southern hemisphere in 2016
  is an explanation to the strong decrease in the southern polar field determined by WSO as compared to our simulations.
  We thus suggest that, while the polar field dataset by WSO is long and relatively homogeneous, the short-term variations might be
  very sensitive to high gradients.

Figure \ref{fig:DM2016} shows the comparison of the evolution of the axial dipole moment between observations and prediction.
The black solid curve in the left panel shows the results based on SDO/HMI synoptic magnetograms, the right panel shows the results
based on NSO/SOLIS synoptic magnetograms. In contrast to the polar field evolution, the averaged values of the 50 random realizations
of the predicted results in colored curves follow the observed ones during 2016-2017, especially for the SDO/HMI maps. The dashed curve in the left panel is axial dipole moment based on SOHO/MDI. According to \cite{Liu2012}, the line-of-sight magnetic signal inferred from the calibrated MDI data is greater than that derived from the HMI data by a factor of 1.40. Here without the factor 1.4, the axial dipole moment based on the SOHO/MDI data matches that based on the SDO/HMI data.

\subsection{Predicting large-scale field evolution during 2017-2020}
We now consider the prediction of the large-scale field evolution during 2017-2020, which is expected cover the remainder of Cycle 24.
Figure \ref{fig:pf2017} shows the time evolution of the polar fields. During the coming one year, the north polar field will
increase and then have a tendency to decrease. The south polar field will remain almost constant. The effects of the random emergence
of the sunspot groups in the activity belts on the polar field start to have an effect from about 2018. The uncertainties of polar
field strength then accumulate. The expected values and 2$\sigma$ uncertainty for the radial polar field in 2020 are
$3.40\pm1.90$ G (N) and $-3.64\pm1.68$ G (S) based on HMI data, $4.74\pm2.13$ G (N) and $-3.64\pm1.87$ G (S) based on SOLIS.
The LOS polar fields in 2020 are $0.93\pm0.55$ G and $-1.06\pm0.49$ G (HMI) and $1.29\pm0.61$ G and $-1.06\pm0.55$ G (NSO).
The expected values are close to the values at the end of Cycle 23.

Figure \ref{fig:MagBtf2017} shows the butterfly diagram based on the SDO/HMI synoptic magnetograms before 2017 and one random
  realization of sunspot emergence after 2017. In the northern hemisphere, the plume with positive polarity started from about 2016
  (denoted by a circle) just reaches the polar region in 2017. The further transport of the poleward flux causes an increase of the
  north polar field after 2017. There are two big sunspot groups emerging close to the equator with abnormal polarities in the northern
  hemisphere during 2016. They are AR12529 and AR12585. They dominate the negative plume which started before 2017 in the activity belt.
  When the negative plume reaches the northern pole, it will stop the field increase and cause the subsequent weak decrease of the north
  polar field, as seen in the upper two panel of Figure \ref{fig:pf2017}. We predict that WSO will measure a distinct decrease of the polar
  field, like the southern polar field during 2016 due to the large gradient around 55 degree since the middle of 2017. In the southern
  hemisphere, weak negative and positive flux was alternatively transported from the activity belt to the polar region. Hence the south polar
  field is expected to remain roughly constant during 2017-2018.

Figure \ref{fig:DM2017} shows the time evolution of the axial dipole moment. The expected values slightly increase during 2017-2020.
At 2020, the expected values and 2$\sigma$ uncertainty for the dipole moment are $1.76\pm0.68$ and $2.11\pm0.69$ based on the initial
conditions from SDO/HMI and NSO/SOLIS CR2185 synoptic magnetograms, respectively. The expected value is within the 2$\sigma$ range
predicted by \cite{Cameron2016a}. The axial dipole moment at the end of cycle 23, i.e. 2008.9, is -1.64 G by SOHO/MDI and -1.5G
by NSO/SOLIS, respectively. Since there is a good correlation between the open heliospheric magnetic flux (strongly related to the
axial dipole moment and the polar field) and next cycle strength \cite{Jiang2007,Wang2009,Munoz-Jaramillo2012,Hathaway2015}, we expect
that Cycle 25 will have a similar strength to Cycle 24.

\section{Conclusions and Discussion}
In this paper, we aimed to predict the time evolution of the large-scale magnetic field over the Sun's surface based on the
well-calibrated surface flux transport model. The main methods and results are as follows.
(1) The Monte Carlo simulation of time evolution of sunspot number generated from the  empirical statistics can be used
to predict the solar cycle at a few years into a cycle.
(2) The polar field can be well predicted one year in advance using the SDO/HMI synoptic magnetograms as the initial condition.
(3) The sharp decrease of the southern polar field during 2016-2017 shown by WSO data was not seen in the HMI data.
We suggest that the apparent decrease shown by WSO is caused by the existence of a large gradient of the magnetic field around 55 degree
and the low spatial resolution of WSO observation.
(4) The northern polar field is predicted to increase during 2017, but strong gradients will be present at a latitude near 55 degrees
and the WSO observations might show a notable decrease of the polar field which should not be seen by other telescopes.
(5) The southern polar field is predicted to be stable during 2017-2018. At the end of 2017, the magnetic field in two poles is expected to
balance.
(6) The axial dipole moment at the end of the Cycle 24 is expected to be $1.76\pm0.68$G, which suggests that Cycle 25
will have a similar amplitude as Cycle 24.

\begin{figure}[!htp]
\includegraphics[scale=0.35]{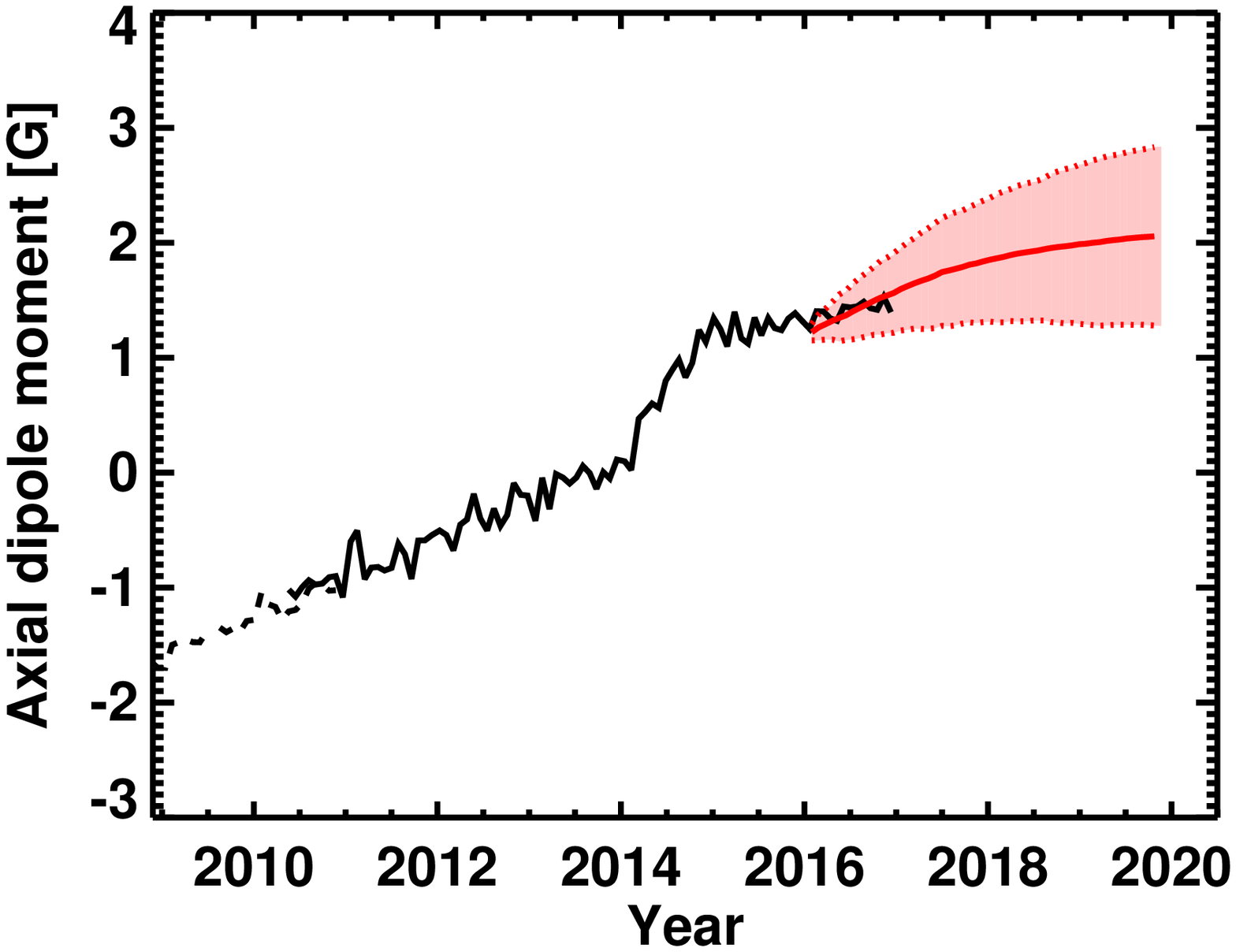}
\includegraphics[scale=0.35]{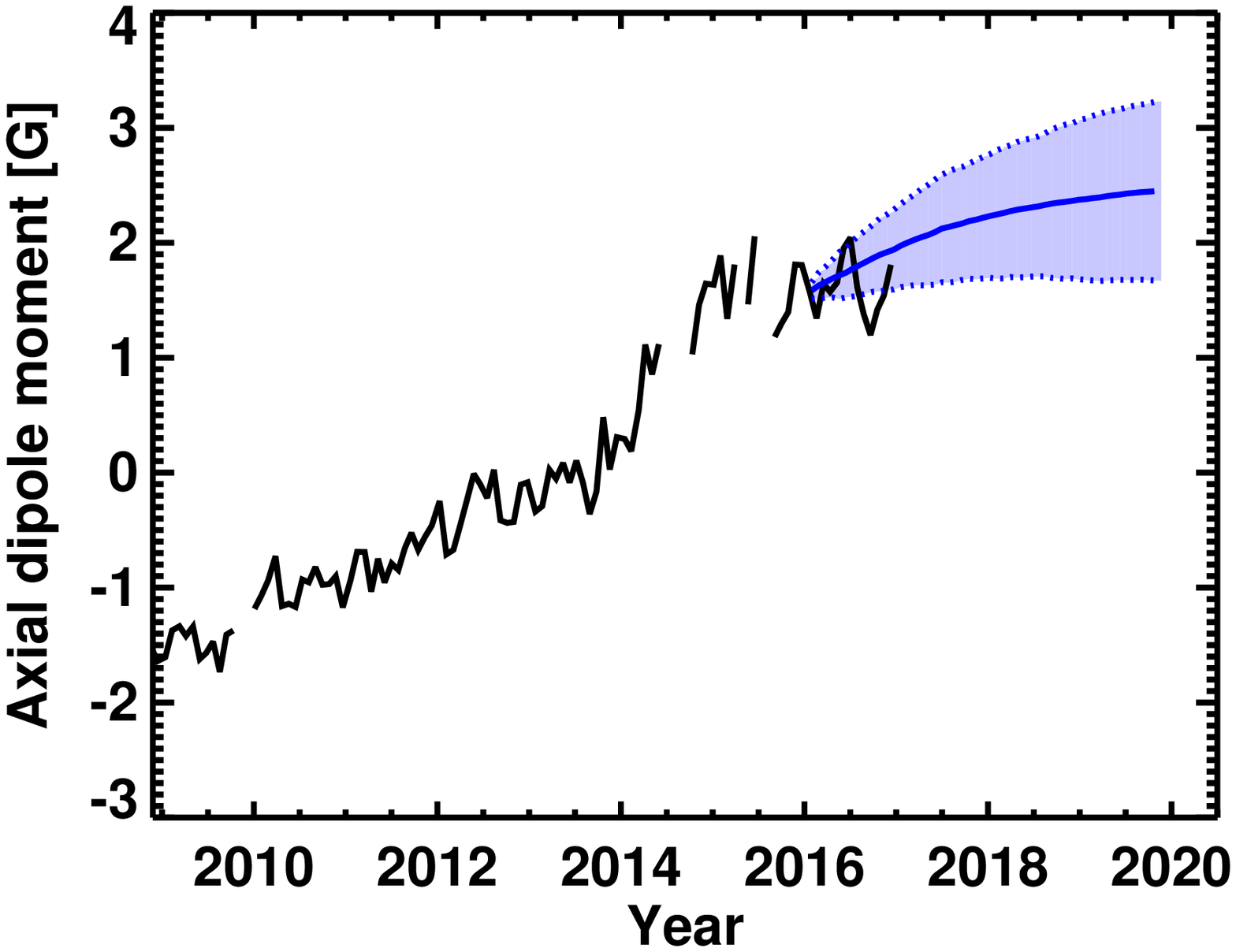}
\caption{Time evolution of the solar axial dipole moment. Black curves: based on SDO/HMI (left panel) and NSO/SOLIS (right panel) synoptic magnetic field data. The red and blue curves correspond to the prediction using SDO/HMI and NSO/SOLIS synoptic magnetogram CR2173 as the initial condition, respectively. Shading region and the central curve show the $\pm2\sigma$ variation and the average of 50 random realizations.}
\label{fig:DM2016}
\end{figure}

\begin{figure}[!htp]
\includegraphics[scale=0.35]{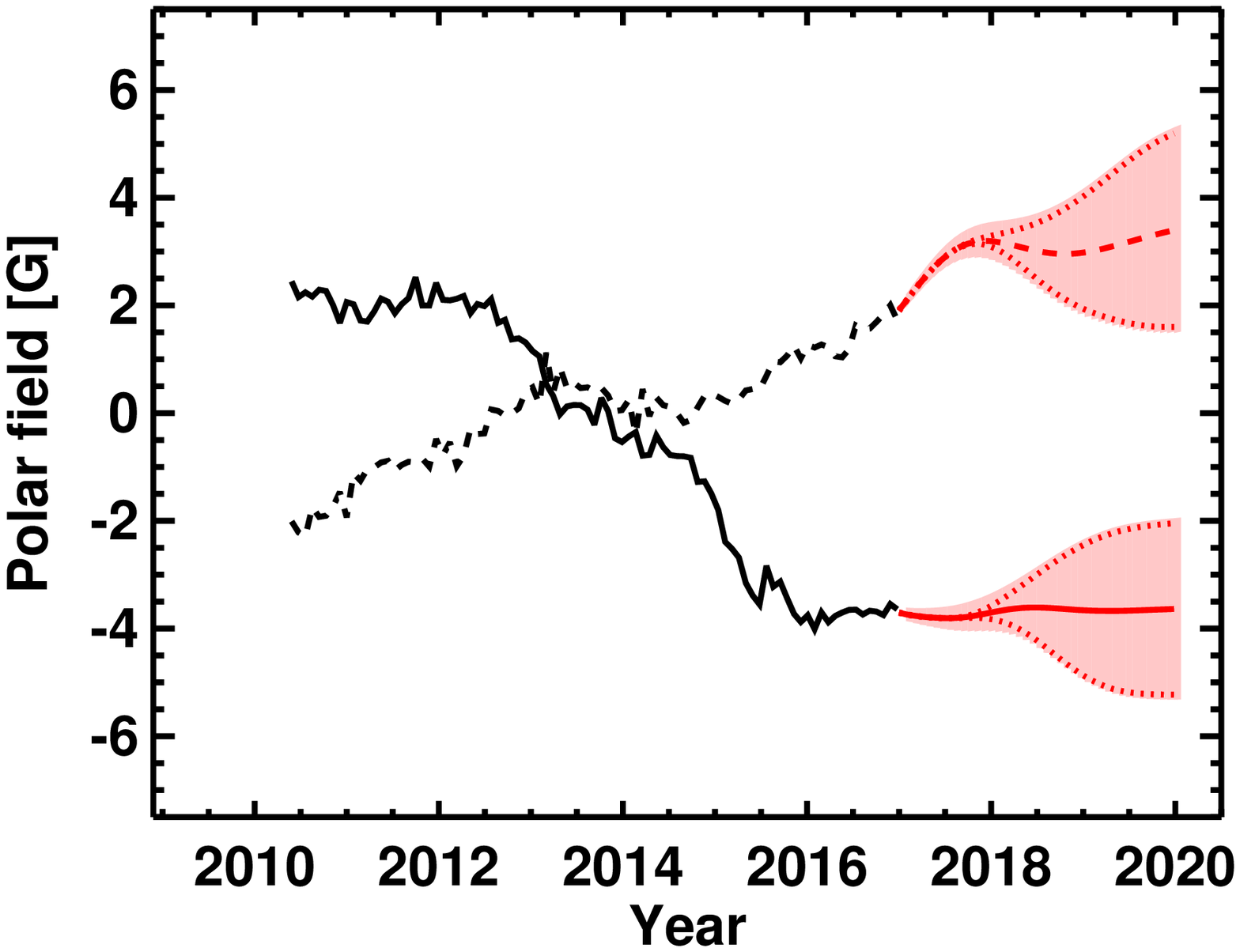}
\includegraphics[scale=0.35]{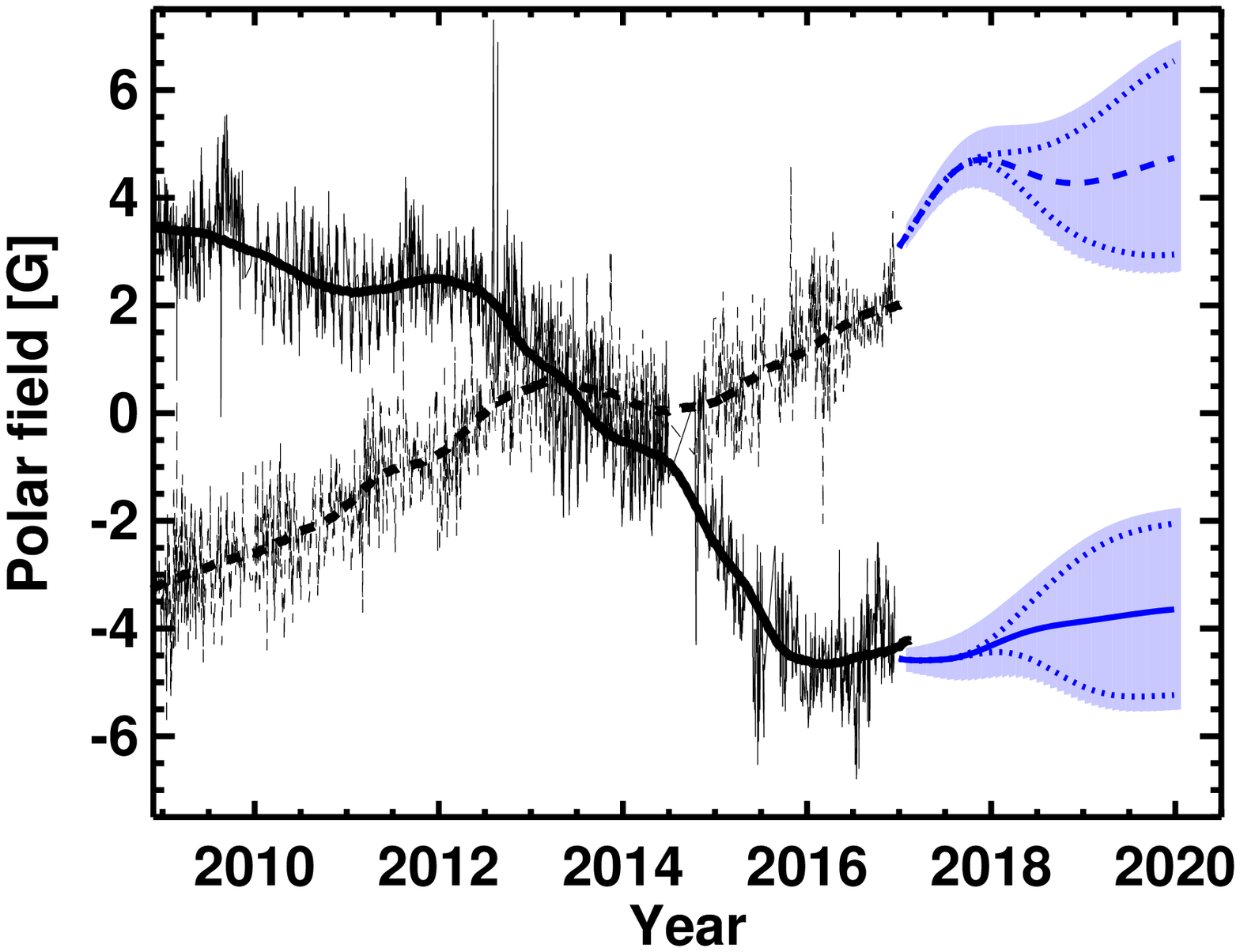}
\includegraphics[scale=0.35]{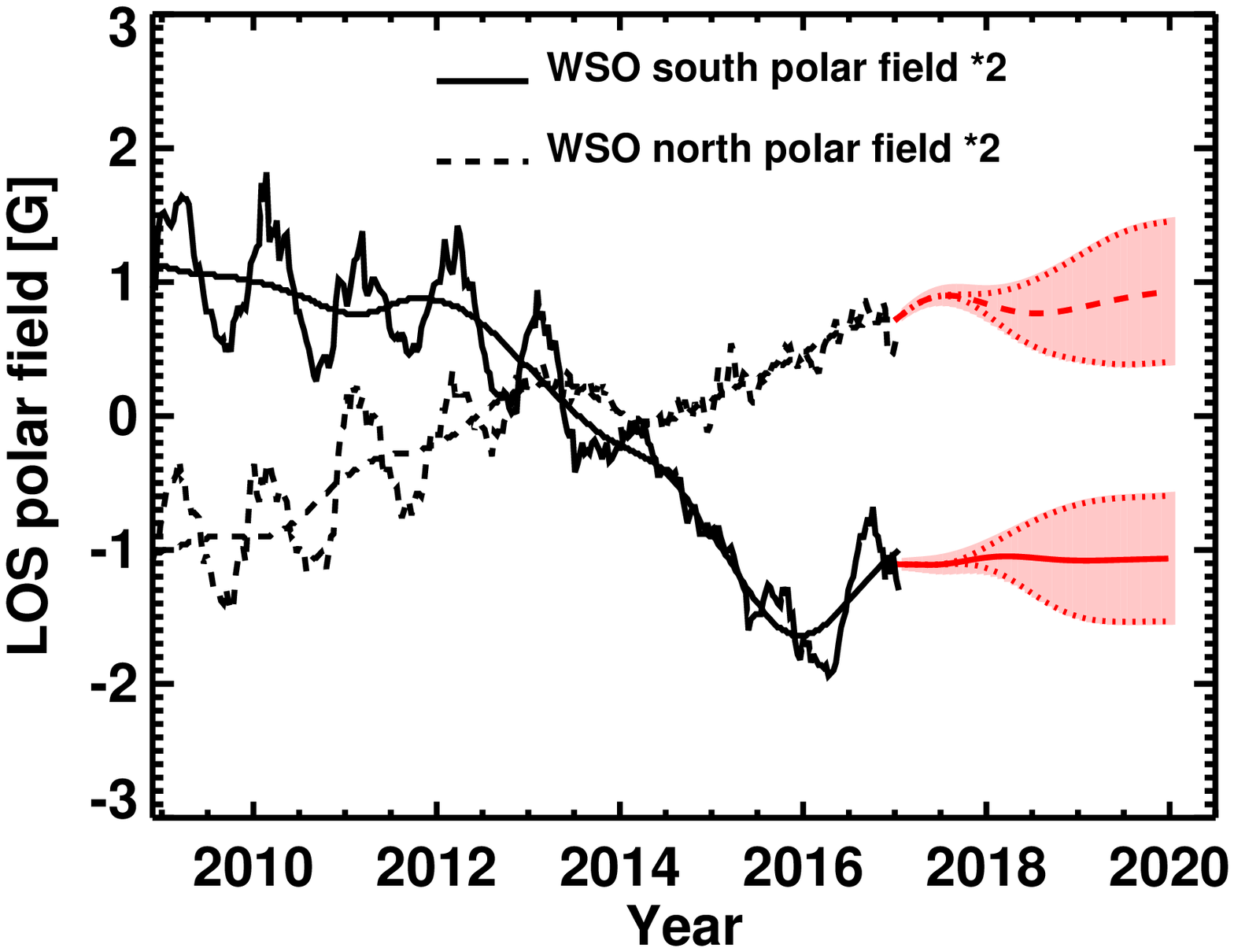}
\includegraphics[scale=0.35]{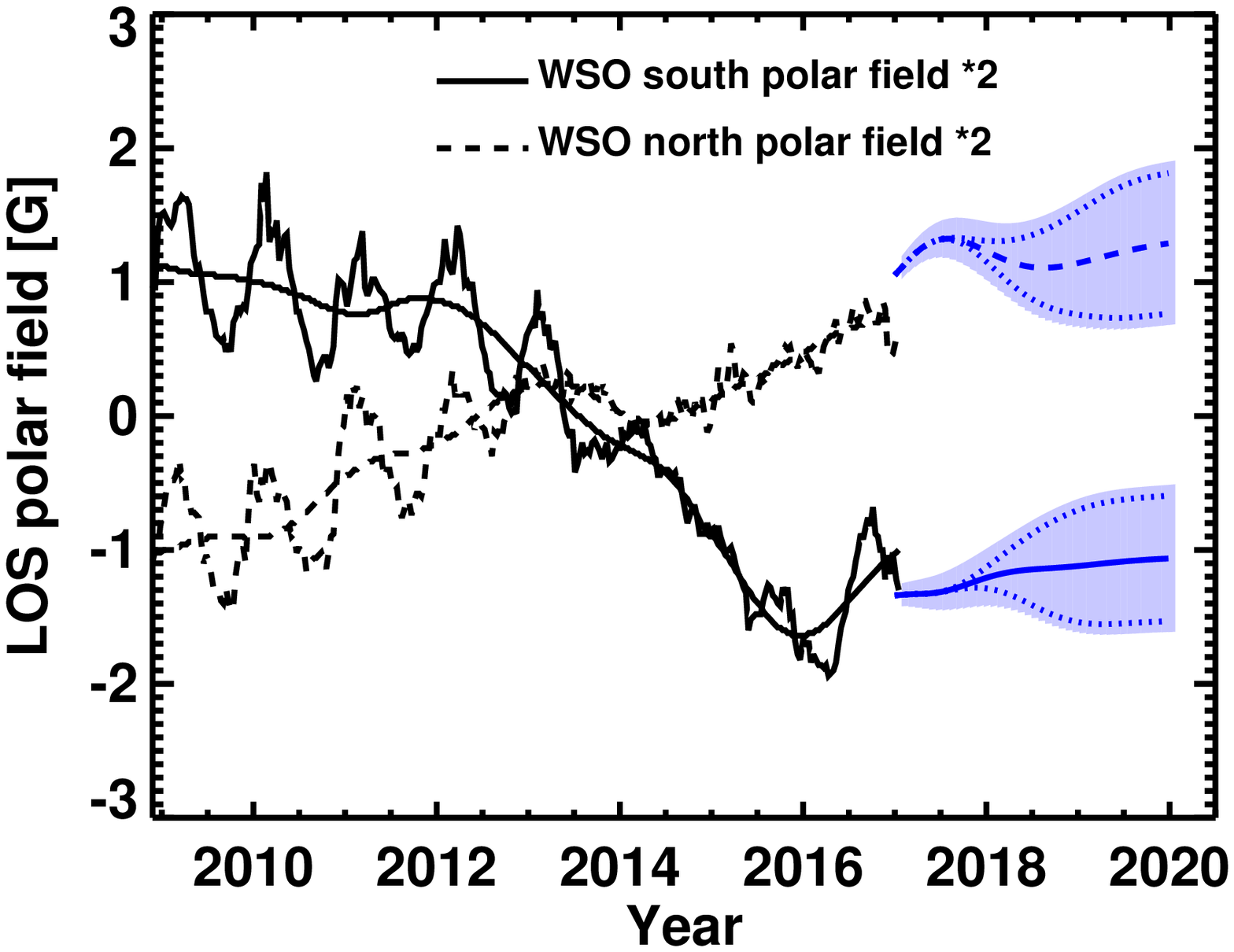}
\caption{Similar to Figure \ref{fig:pf2016}, but for the simulations using the synoptic magnetogram CR2185 as the initial condition.}
\label{fig:pf2017}
\end{figure}

\begin{figure}[!htp]
\includegraphics[scale=0.5]{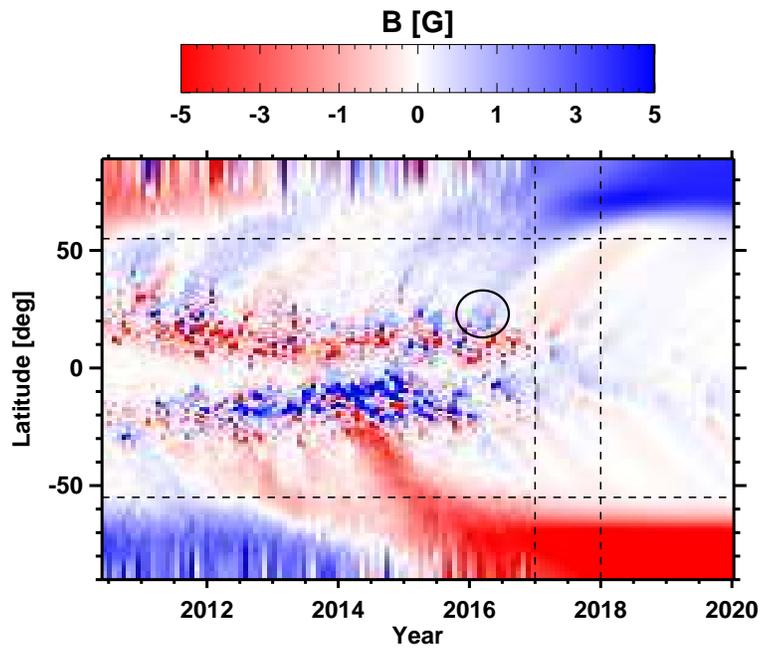}
\caption{Time-latitude diagram of the longitudinally averaged radial magnetic field at the solar surface observed by  SDO/HMI during 2010-2017
  and simulated with one random realization of flux source during 2017-2020.}
\label{fig:MagBtf2017}
\end{figure}

\begin{figure}[!htp]
\includegraphics[scale=0.5]{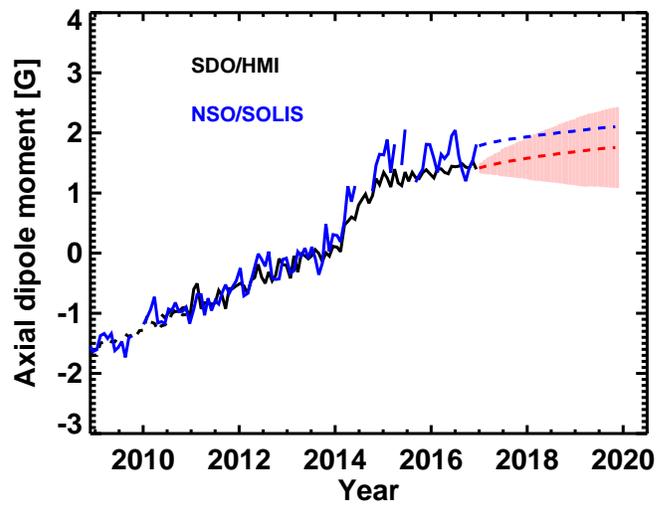}
\caption{Predicted evolution of the axial dipole moment until 2020 based on 50 SFT simulations with random
sources starting from the observed synoptic magnetograms CR2185. Black line indicates
values determined from SDO/HMI. The figure layout corresponds to that of Figure \ref{fig:DM2016}. The result based on
NSO/SOLIS is also presented without the $2\sigma$ shading, whose range is similar to the red one.}
\label{fig:DM2017}
\end{figure}

\cite{Hathaway2016} also used a SFT model to predict the polar field and the axial dipole strength.
They listed the primary differences between their methods and \cite{Cameron2016a}'s in three points.
Here we give some comments on the differences.
Their first point concerns that an approximation of the convective motion as a scalar diffusivity is used or not.
Recently, \cite{Martin-Belda2016} demonstrated that on the scales we are interested in, the flux dispersal due to turbulent flows can
be described as a diffusion process. Figures 4 \& 6 in \cite{Hathaway2016} show that variations in the convective flow pattern
have minor effects on the polar and axial dipole fields. Hence the explicit modelling of the convective flow pattern in
the SFT code has negligible effects on the improvement of the results.
(2) Sunspot emergence in previous solar cycles is used as a representation of the sunspot emergence of future cycles in
\cite{Hathaway2016}. This is feasible for the declining phase since all the cycles show the similar decline phase \cite{Cameron2016b}.
But if we hope to predict the sunspot emergence and the polar field evolution at the early state of a cycle, the Monte Carlo random
realizations based on statistical relations is more flexible.
(3) Whether the uncertainty in the initial conditions is considered or not. As we have seen in Section \ref{sec:results},
the observed magnetograms significantly affect the initial and subsequent polar field and axial dipole field. Moreover, a constant
projection factor was applied to WSO LOS field to compare with the radial field by \cite{Hathaway2016}. Our LOS polar field is
obtained based on Equation \ref{eq:polar1}, which includes a latitudinal dependence of the projection factor. Our polar field evolution
is consistent with HMI results, but failed to reproduce the significant decrease of southern polar field generated by WSO.
\cite{Hathaway2016} only compared their results with WSO and they managed to predict the decrease of the polar field measured by WSO
(see their Figures 6 \& 7).
The axial dipole strength is expected to keep increasing if there are no sunspots emergence having abnormal polarities.
We see the increase of the average axial dipole field in Figures \ref{fig:DM2016}. Only a few of the random realizations shows
a decrease in the axial dipole field (in red shades). In contrast, all the random realizations in \cite{Hathaway2016} show a
decrease of the axial dipole strength since about 2018.
The causes of these differences between our results and those in \cite{Hathaway2016} are not immediately clear.

Acknowledgments
We are deeply indebted to Dr. Robert Cameron for the language corrections of the manuscript. We are grateful to the two anonymous referees for constructive suggestions. The SDO/HMI data are courtesy of NASA and the SDO/HMI team. Sunspot data are from the World Data Center SILSO, Royal Observatory of Belgium, Brussels. NSO/SOLIS data are courtesy of NISP/NSO/AURA/NSF. This work is supported by the National Natural Science Foundations of China grant 11522325.

\end{document}